\documentstyle{l-aa}

\begin{document}

\thesaurus{06(08.01.1; 08.01.2; 08.02.1;  08.12.1; 08.18.1;  08.05.3}

\title{The age-mass relation for chromospherically active
binaries}

\subtitle{III. Lithium depletion in giant components\thanks{Based
on
 observations collected with the 2.2m telescope of the
German-Spanish 
 Observatorio de Calar Alto (Almer\'{\i}a, Spain), and with the
2.56m Nordic 
 Optical Telescope in the Spanish Observatorio del Roque de los
Muchachos of 
  the Instituto de Astrof\'\i sica de Canarias (La Palma, Spain)}}

\author{D.~Barrado y Navascu\'es \inst{1,2} 
         \and  E.~De~Castro \inst{3}
         \and M.J.~Fern\'andez-Figueroa \inst{3}
	 \and M. Cornide \inst{3}
         \and R.J.~Garc\'{\i}a L\'opez \inst{4}
         }

\offprints{D.~Barrado y Navascu\'es, dbarrado@cfa.harvard.edu}

\institute{MEC/Fulbright Fellow at the Smithsonian Astrophysical
Observatory, 
        60 Garden St, Cambridge, MA 02138, USA.
        \and
           Real Colegio Complutense at Harvard University,
        Trowbridge St, Cambridge, MA 02138, USA.
        \and
        Dpto. Astrof\'{\i}sica. Facultad de F\'{\i}sicas.
           Universidad Complutense. 
           E-28040 Madrid, Spain.   
        \and
           Instituto de Astrof\'\i sica de Canarias. E-38200 La Laguna, 
           Tenerife, Spain.}                     
  
\date{Received date, ; accepted date, }

\maketitle

\begin{abstract} 

We present a study of the lithium abundances of a sample of evolved components
of Chromospherically Active Binary Systems. We show that a significant part of
them have lithium excesses, independently of their mass and evolutionary stage.
Therefore,
 it can be concluded that Li abundance does not depend on age for giant
components of CABS.
These overabundances appear to be closely related to the stellar rotation, and
we interpret them as a consequence of the transfer of angular momentum from
the orbit to the rotation as the stars evolve in and off the Main Sequence, in
a similar way as it happens in the dwarf components of the same  systems 
%[Barrado et al. 1997, A\&A 326, 788]
 and in the Tidally Locked Binaries belonging to the Hyades and M67.
%[Barrado y Navascu\'es \& Stauffer 1996, ApJ 475, 313].

\keywords{stars: abundances -- stars: activity -- stars: binaries: close --  
stars: late-type -- stars: rotation  -- stars: evolution}

\end{abstract}

\section{Introduction}

This is the third paper of a series devoted to the study of the stellar
properties of Chromospheric Active Binary Systems (CABS). This heterogeneous
group of stars includes systems which have evolved components together with
binaries with both members inside the Main Sequence (MS). Strassmeier et al.
(1993) presented a characterization of this type of binaries and a catalog
complete up to that date. In Paper I (Barrado et al. 1994), we studied the
evolutionary status of CABS using the effective temperature ($T_{\rm
eff}$)--Radii plane and estimated their ages by comparing masses, radii and 
temperatures for both components with theoretical isochrones. The main
conclusion of that paper is that  CABS are in a particular stage of their
stellar evolution. The high stellar activity appears in CABS when one of the
components is evolving off (or has already done so) the MS, crossing the
Terminal Age Main Sequence (TAMS). Classifying the systems on the $T_{\rm
eff}$--Radii plane, we established in Paper I that CABS appear divided in three
different groups: MS stars with masses $<$1.7 M$_\odot$, evolved stars with
masses close to 1.4 M$_\odot$, and giant stars having masses in the interval
2.5--5.0 M$_\odot$. The second group could be subdivided into other three
types: subgiants evolving off the MS, subgiants at the bottom of the Red Giant
Branch (RGB), and giants ascending the RGB above the previous
sub-classification. However, not all CABS have accurate values of radii and
masses making so impossible their location in any of these groups using these
criteria. For this reason, Paper II (Barrado y Navascu\'es et al. 1997) used
other strategy to classify the components of CABS. We studied the evolutionary
status of the whole sample of CABS listed in Strassmeier et al. (1993) using
the photometry. Visual magnitudes, colors, and individual distances (when
available), allowed us to locate the stars in a Color--Magnitude Diagram (CMD).
CABS are concentrated in specific regions: MS stars, stars evolving off the MS,
those at the bottom  and at the top part of the RGB (B--RGB and T--RGB,
respectively) and components located in what we called 
 the horizontal branch (HB) due to its geometrical configuration. Note that 
our nomenclature does not refer to the real horizontal branch, formed
by RR Lyrae and other types of stars. The CABS have masses around 5 
M$_\odot$ and are first crossing the giant gap or are in a later evolutionary
stage, and their position in the Color-Magnitude Diagram does not allow 
to establish the evolutionary status (see Figure 1). To avoid confusion, 
we have named them here high luminosity (H-Lum) stars
 (compared with the other stars in this sample).

Paper II dealt also with the lithium abundances of those components on the MS
or just evolving off it. We concluded that these stars have a Li excess, which
could be related to their binary nature. In particular, we interpreted the
excesses in the context of the transport of angular momentum between the orbit
and the stars, which would inhibit mixing mechanism inside and lead to
inhibition of Li depletion. Other scenarios, such as the depletion induced
by internal gravity waves and the role of magnetic fields in inhibiting this
process, can also be accommodated to the observations. We refer the reader to
Paper I and Paper II  for a detailed description of the properties of  CABS and
for a introduction to the Li depletion phenomenon among them. Additional
information can be found in Fern\'andez--Figueroa et al. (1993), Barrado y
Navascu\'es (1996) and Barrado y Navascu\'es (1997).

In this work (Paper III), we study the Li abundances of evolved components of
CABS, which provide complementary information to that obtained from dwarfs. In
Section 2, we describe the data and the evolutionary status of our sample of
stars, whereas Sections 3,  and 4 contain the analysis and discussion of the
lithium abundances. The conclusions are summarized in Section 5.

%\section{Giant components of CABS}

\section{Program stars and observations}

The basic data --namely, photometry, stellar masses, radii, distances, orbital 
and rotational periods-- were selected from the Catalog of CABS 
(Strassmeier et al. 1993 and references therein). 
However, we revised the photometry for a significant fraction of the sample, 
carrying out a deconvolution of the photometry in order to locate each
component in the Color-Magnitude Diagram and to compute accurate temperatures.
Essentially, we used the available information to obtain colors and magnitudes
for each component.
Details about the process can be found in the footnotes of Table 1 and in Paper
II. A more exhaustive study can be found in Barrado y Navascu\'es (1996).
Table 1 shows the systems 
which have been studied in this paper: column 1 lists the name of the system, 
column 2 the HD number, columns 3, 4, 5 and 6 the (U--B), (B--V), (V--R)$_{\rm
J}$ and (V--I)$_{\rm J}$ colors, respectively, for each  component. Column 7
lists the distance, whereas the 8$^{\rm th}$ and 9$^{\rm th}$ columns contain 
the apparent visual magnitude and the absolute magnitude computed with the 
previous 2 columns, respectively. In this last case, we also took into account
other  information in order to obtain the most accurate estimate of the
magnitude of  each component in the binary system (see notes in Table 1). 
Finally, column 10 declares if the system is SB1 or SB2, column 11  shows our
evolutionary classification: the bottom or the top part of the
Red Giant Branch (B--RGB and T--RGB), and those stars with higher luminosity
(H--Lum). This last case corresponds to components of CABS which are crossing
for the first time the giant gap or are already burning helium, but due to 
their particular location in the CMD, it is not possible to determine the
exact evolutionary stage. Column 12  provides
the spectral type. The last column contains notes about the method used  to
compute the absolute visual magnitude and the photometric indices for  each
component (e.g., deconvolving the photometry. See Barrado y Navascu\'es \&
Stauffer 1996). Note that the distances listed in Table 1 are pre--{\it
Hipparcos}. We have preferred to maintain the previous values for coherence
with Paper I and II. A revision of these results, together with  new Li
abundances derived from spectral synthesis, is in progress. In total, our
sample has 70 systems, which contain 76 evolved components. 
A Color--Magnitude Diagram can be found in Figure 1. Note that we have 
used our deconvolved values of the photometry.

The observations were carried out in two different observing campaigns at the
coud\'e focus of the 2.2m telescope of the Calar Alto Spanish--German
Observatory  (June 1993 and May 1994) and in an additional run at the 2.56m
Nordic Optical Telescope of the Spanish Roque de los Muchachos Observatory
(September 1994). In the first case, we used the Boller\&Chivens spectrograph
at the Coud\'e focus. For the last campaign, IACUB was utilized. This is a
echelle spectrograph without cross dispersion. Since we used a filter
to select the order around 6700 \AA, no scatter light belonging to other 
order polluted our spectra.

 Spectral resolutions ranged from 30,000 to 50,000 at
$\lambda$ 6700 \AA. The spectra were reduced using standard procedures and the
MIDAS\footnote{ Munich Image Data Analysis System is a program developed by
European Southern Observatory, Garching.} package. More details can
be found in  Paper II. Figure 2 shows a handful of our spectra. It can be 
appreciated their high quality. The signal-to-noise ratios
range from 100 to 160.

\section{Temperatures, lithium abundances, masses and ages}

\subsection{Effective temperature}

The determination of the lithium abundances is strongly sensitive
to the effective temperature. Therefore, we made real effort to try to 
derive them as accurately as possible
 (taking into account that these stars are binaries, 
and the observed photometry is the result of the combined light from 
both components). We used different color indices --namely, 
(B--V), (V--R), and (R-I), and computed the (V--I) using the last two
 indices--. 
%
%However, there is an element of subjectivity in this process.
%
First, we select the whole sample of CABS published in the Strassmeier et al.
(1993) catalog. The deconvolution of the photometry was performed for the
whole sample. Then, we computed different temperatures using the colors and
the spectral type. The comparison between these values shows that the 
differences are quite small.
In the case of the sample studied in this paper, 
the average temperature derived from
 (V--I)$_{\rm J}$ and (R--I)$_{\rm J}$ was selected for the cool component
of each system, if these indices are known
(temperature scales from Cayrel et al. 1985 and Carney  1983, respectively).
For the hot components we used the calibration by B\"ohm-Vitense (1981),
based on (B--V). Although it depends on the availability of the data of
different colors, in general the borderline is about 5000 K. We believe 
that this is the most accurate method to derive temperatures, since the 
contamination of light from the primary (hot component) is less in the red 
filters than in  (B--V).
 The spectral type and the Schmidt--Kaler (1982) scale were
used for those components which do not have photometry accurate enough
and to check the previous values.  Results
are listed in Table 2. The comparison between temperatures estimated using
different calibrations does not show any relevant bias which could affect
significantly the discussion performed in this work. The average value
of the differences in temperatures (from the three color indices and the
spectral type) is 110 K. Usual names, HD numbers,
effective temperatures and radii, equivalent widths --EW(Li$+$Fe) and
EW(Fe) as measured in this work and by Pallavicini et al. (1992)--,
 the final values of the  lithium equivalent width --EW(Li)--,
 and comments can be found in this table, as well as the
continuum  correction factors (CCF). These last values take into account the
contribution to the flux of each component and are very important  to derive
the actual equivalent widths. The CCF were determined using radii and 
temperatures of both components or, if the first ones were not available,  from
the visual magnitudes. We have also included all CABS studied by
Pallavicini et al. (1992) and Randich et al. (1993, 1994) which were not 
observed during our observing runs.

\subsection{Lithium abundance}

The Li\,{\sc i} doublet at $\lambda$ 6708 \AA\ is in fact a complex
feature. At our resolutions, it appears as a single line and it could be
blended with other spectral characteristics, either from the same star (like
Fe\,{\sc i} $\lambda$ 6707.4 \AA) or from the other component, since an
important part of our systems are SB2. We measured the Li equivalent widths
(EW) by fitting Gaussian curves to the relevant features, taking into account
the phase information. Details of the procedure can be found in Paper II.
Measured EW were corrected for the contribution to the continuum by the
other component.

Table 3 lists the corrected lithium equivalent widths derived using the 
observed values and the CCF. It also includes four different lithium 
abundances: the values derived in this work, and those published by Pallavicini
et al. (1992), and Randich et al. (1993; 1994). Other stellar parameters,
such as masses, orbital and rotational periods,  and an estimation
of the age, appear in the last columns.

The final abundances were derived using the
curves of growth of (CoG)
 Pallavicini et al. (1987). In order to reach lower temperatures, we added
the CoG by  Soderblom et al. (1993) corresponding to temperatures of
4000 and 4250 K and a gravity of Log g=4.5. 
To obtain abundances in the same way for the whole sample,
using our temperatures and these curves of growth, 
we had to retrieve lithium equivalent widths from the published abundances 
by Randich et al. (1993; 1994). Since the comparison between the stars in
common between these works and Pallavicini et al. (1992) shows that
they are very similar, and this last study used the same CoG as us, we proceeded
in the following way. First, using Randich et al. abundances and temperatures,
we computed the equivalent widths. Second, with these values and our own
temperatures, recalculated the abundances. As can be seen from the 
comparison of columns 5th, 6th, 7th and 8th, the abundances derived in 
most of the cases are quite similar. Essentially, the important departures
appear when the temperatures selected by Pallavicini et al. (1992) and
Randich et al. (1993; 1994) were very different from our values. In these 
cases, our temperatures are normally lower and, therefore, the abundances
tend to be smaller.  

  The final errors in the Li abundances can be
estimated as $\sim$0.4 dex (see Paper II for details). 
In this case, there is an additional source of error for the abundances:
the gravity. Pallavicini et al. CoG were computed for 
evolved and main sequence stars (we selected the appropriate set for each 
star), whereas Soderblom et al. CoG were obtained for main sequence
stars.  However,  the comparison of the derived abundances for the same 
equivalent widths and temperatures and different gravity shows that the 
differences are negligible ($\sim$0.02) and they do not contribute
in a significant way to the final errors.

\subsection{Masses and ages}

We selected the masses from the catalog of chromospheric active binaries
by Strassmeier et al. (1993). In some cases, only the mass function has
been published, but a value of the inclination has been obtained and it is
possible to compute the mass. These cases are denoted with the 
flag ``Msin$^3$i'' in column 9 of Table 3. Unfortunately, about half of the
stars in this sample have unknown masses. The location of the component
on the Radius-Teff plane and Color-Magnitude Diagram allows an estimate of
the mass. For this goal, we used the Schaller et al. (1992) and Schaerer 
et al. (1993a,b) evolutionary tracks. The flags ``RT'' and ``CM'' indicate
those masses derived by this way. It should be kept in mind that this is a
crude estimation of the masses, since there are several factors that 
contribute to the uncertainty in the process. Among others, the errors in the
parameters we used (temperature, radii, visual magnitude, color, distance),
the set of evolutionary tracks (since different metallicity shifts the
positions and, therefore, a different mass is derived).  In order to estimate 
the uncertainties of the 
the computed values, we have proceeded in the same fashion for those
stars with known masses. The comparison between these values (the real
ones and those from the CMD) shows that in general the agreement is good 
enough for low mass stars, in the range 1-3 M$_\odot$. However, more massive
stars, as derived from the CMD, show, in general, larger values than the 
dynamic masses. The effect is more dramatic for those stars with masses
around 5 M$_\odot$ (CMD masses). An estimation of the error in the mass
for this range is 2 M$_\odot$. For the low mass, it is about 20\%.
Thus, the high masses quoted in Table 3 with the flag ``CM'' should
be taken with some caveats, since they are only intended as indicative 
value.

 Ages were estimated
following Paper I. We showed there that  there is a clear correlation between
mass and age for CABS. This relation can be used to estimate the age of those
components whose radius is not known. In several cases, we assigned the
age computed for the more massive component to the secondary, since the primary
is the component which determines the evolutionary status of the system
(the enhanced chromospheric emission arises normally from it). On the other 
hand, there are several systems which show clearly an anomalous evolution, 
with evidences of a significant transfer of material from one component 
to the other (e.g., a giant primary --brighter- with a main sequence as a
secondary, having  inverted mass ratios). 
Due to the complexity of the evolution of these systems, no age was estimated.

The  actual expressions for evolved
systems are:

\begin{equation}
\log~{\rm Age} = (9.81\pm0.04) - (2.64\pm0.10)\log~({\rm Mass}/{\rm M}_\odot),
\,\,{\rm luminosity~ class~ III.}
\end{equation}

\begin{equation}
\log~{\rm Age} = (10.00\pm0.02) - 
(3.25\pm0.10)~\log~({\rm Mass}/{\rm M}_\odot),
\,\,{\rm luminosity~ class~ IV.}
\end{equation}

\noindent An illustration of these relations can be seen in 
Figure 3 of Paper I.

\section{The dependence of lithium on several stellar parameters}

\subsection{Li abundance vs. effective temperature}

In Figure 3a,b,c, we plot Li abundance against effective temperature. 
%
%%%%%
%For comparison,
%a solid line indicating the Li abundances of Hyades stars is also shown (note
%that all stars in this last group are in the MS).
%%%%
%
 As it happens in  members of open clusters, there is a clear trend between
both quantities: the hotter the star, the larger the abundance. However, there
is an important scatter and most of the evolved components of CABS have
temperatures in the range 5000--4300 K and abundances log N(Li)$\sim0.5-1.2$
(in the customary scale where log N(H)=12). Similar plots for CABS have been
produced recently by different authors (Pallavicini et al. 1992;
Fern\'andez-Figueroa et al. 1993; Randich et al. 1993, 1994), who concluded
that a significant number of these stars  have Li excesses, although it is not
a characteristic as a group.  
This dependence between lithium abundance on effective temperature is
due to the fact
 that evolved components of CABS have changed their  atmospheric
physical parameters due to the evolution off the MS, cooling down due to the
expansion of the external layers. A proper comparison between these stars and
MS stars which are members of open clusters requires the knowledge of the
stellar mass. As can be seen in Figure 3a,
 for the same temperature, there are stars which have very
different mass (e.g., from 0.29 M$_\odot$ to 5 M$_\odot$). 
Although these stars
are in similar position in the $T_{\rm eff}$-log~N(Li) plane, the mechanisms
responsible for the Li depletion are completely different. Stars  with masses
larger than $\sim 1.5$ M$_\odot$, due to their shallow convective envelope,
have not destroyed any photospheric Li during the MS life--time, whereas stars
having masses in the range 0.8--0.9 M$_\odot$ have burnt at least 99\% of their
original Li content at the age of M67  {\it before} they have left the MS
(Chaboyer et al. 1995; Balachandran 1995; Barrado y Navascu\'es et al. 1998).
All of them, after they leave the MS phase, experiment a diminution of their
photospheric Li due to the dilution process (see below), when the external
layers are mixed with internal ones, poor in Li. Other mechanisms, such as
turbulent circulation, could also be present at this stage.
In addition, some CABS, with low orbital periods,  have undergone a process
 of transfer of material when the primary has evolved
off the MS.

Figure 3b contains comparison between our sample of CABS and giants
 belonging to open clusters (Gilroy 1989). The binaries are shown as 
solid symbols and the cluster giants appear as open circles. The ages
of the clusters range from 5$\times10^7$ to 2$\times10^9$ yr. No obvious 
difference can be concluded from this plot. For the area in common between
both groups, the distribution of lithium abundances is quite similar and
any difference could be attribute to the differences in the analysis and/or
to the uncertainties of temperatures and abundances. A more meaningful
comparison is carried out in Figure 3c. We selected a those CABS having
masses in the range 1.1$\le$Mass$\le$1.5 M$_\odot$ (solid symbols). The open
 circles represent single, evolved M67 stars (Barrado y Navascu\'es et al.
1998). The lithium abundances in both samples were derived in similar
way, using the same set of CoG. The M67 members have masses around 1.3 
M$_\odot$, as derived from the CMD.
 Most of these stars have only upper limits for the lithium
abundances. The visual inspection of the figure indicates that the abundances
are systematically larger in the case of the CABS (half of them have
an abundance about 0.6 dex larger than the values of the M67 members for the
same temperature). However, it should be kept in mind that the masses of the 
CABS cover a larger range and that the M67 giants could come from MS stars
in the Li gap.

\subsection{The mass--log~N(Li) plane}

Knowledge of the stellar masses of evolved components of CABS allows us to
avoid the evolutionary effects when comparing Li abundances. Figure 4 shows Li
abundance against mass. Components at the bottom of the RGB are represented as
open circles, whereas  those at the top of the RGB appear as solid squares, and
giants belonging to the H--Lum are shown as open triangles. The average Li
abundance of single Hyades stars and the predicted abundances for subgiant
and giant stars due to dilution (Iben 1966, 1967a,b; after Scalo \& Miller 
1980) are   included as a solid lines.
 This process was described
originally by Iben (1965) and appears when a star evolves off the MS.
When it crosses the giant gap and climbs the RGB, the convective envelope
increases its fraction in mass. Material poor in Li arises to the surface and
is mixed with Li-rich material. This process stops when the star achieves the
maximum size of its convective envelope, but the depletion of the surface Li
can continue if there is a further mixing related to rotation (Ryan \&
Deliyannis 1995).  With the
exception of three stars, the rest of the sample objects have abundances at
least 1 dex lower than the  ``cosmic'' value
(dashed line at log~N(Li)=3.2; Mart\'{\i}n et al.
1994), indicating that they have suffered Li depletion and/or dilution
(in this last case, as the start to ascent the RGB). One of these three
stars is in the middle of the Li gap (Boesgaard \& Tripicco 1986a,b) and two of
them (HR5110 and $\alpha$ Aur) have signs of an anomalous evolution (see Paper
I) and/or seem to be first crossing the giant gap.

Several stars have lithium abundances between the cosmic value and the 
values predicted by dilution. However, they are either massive stars
(H--Lum or T--RGB) in an ambiguous evolutionary stage or stars classified
as B--RGB, but at the beginning of the evolution in the RGB. Therefore, 
their high lithium abundances could be explained as a result of a limited
lithium dilution, since the convective envelope has just started to 
expand inwards. However, other mechanisms could be acting (see below). 

Two thirds of the stars in the sample have
abundances below to the dilution limit. However, this is a maximum value, 
since MS experiment lithium depletion before they evolve off it, and a further
depletion could take place later. Brown et al. (1989) showed that $\sim$99\%
of the large sample of giants 
studied by them have a lithium abundance well below
the predicted abundances due to the mixing process during the first
dredge-up. In particular, the average abundance is 
$<$Log N(Li)$>_{\rm giant}\le$0.0. However, less than a dozen components
of CABS have similar abundances to the field giants, and most of the stars 
in our sample have values clearly above them.

On the other hand,  when the comparison is made for stars in the same mass
range, some T-RGB stars have abundances higher than those characteristic of
B--RGB, which are in a evolutionary stage previous to that corresponding to
T--RGB. A similar situation appears if the comparison is made between T--RGB
and H--Lum stars (but in this case, some H-Lum could be crossing the giant
gap for the first time). Since evolved stars become older as they climb the
 RGB or the 
asymptotic giant branch (AGB), it seems that part of the evolved stars in our
sample have a Li abundance which does not depend on age or evolutionary stage.
This conclusion, derived from a comparison of abundances, masses and luminosity
classes, is the same as that one obtained from the study of ages (see next
section).

The scatter of massive stars ($\sim$5 M$_\odot$) is remarkable: the Li
abundances vary in 2.5 dex. Brown et al. (1989), using a large sample of giant
stars, showed that only about 2\% among  them  have Li abundances higher than
log~N(Li)=1.5 and a 7.6\% larger than 1.0 dex, in good agreement with the
theoretical predictions based on the dilution. However, the lithium abundances
of the massive stars in our sample follow a complete different pattern. Since
these stars seem to
 have the similar masses (the uncertainties are large for most
of them) and are in the same evolutionary stage, it seems that some have
retained the lithium at the maximum level predicted by dilution, whereas others
have depleted it (the maximum difference is 1.8 dex). Note that almost
all of them have abundances larger than the stars studied by 
Brown et al. (1989).
 Sackmann \&  Boothroyd (1995) have shown that circulation
processes in AGB stars can transport material and can explain the high Li
abundances and the  low $^{12}$C/$^{13}$C ratios found in these stars
(a low  $^{12}$C/$^{13}$C ratio indicates a deep mixing).  These abundances
would be produced in layers at high temperatures close to the core in stars
with masses in the range 4--7 M$_\odot$ in the so called ``hot bottom
burning'', and would be transported to the surface during the second
dredge--up via deep circulation. Moreover, this transport would be effective
also for stars in the RGB, with inferior masses. Li would be created via the
Cameron--Fowler mechanism (``cool bottom processing'') in the layer where H is
burned. In this mechanism, Be would be first synthesized  due to interactions
between $^3$He and $^4$He. This new material  would be rapidly transported
toward the exterior, due to deep circulation,
 where Li would be created by electronic capture. However,
this scenario does not explain the large scatter in the Li abundances of stars
in the same evolutionary stage and should be acting in normal stars.
Only if the deep circulation responsible of the transport of the Be to the 
external layers would depend on rotation, this mechanism could be used
as an explanation for the lithium excesses.
  Fekel \& Balachandran (1993) established that
single giants with strong stellar activity have large Li abundances and
suggested that these abundances could be related to transport of angular
momentum and material from the interior to the surface. 
They have interpreted that, when these stars
expand their convective envelopes inwards while ascending the RGB, they
transport angular momentum toward the surface and new synthesized Li from the
nuclear reactions $^3$He($\alpha$,$\gamma$)$^7$Be($e$,$\nu$)$^7$Li (Simon \&
Drake 1989). This scenario is fraught with problems as well. It
presupposes a reservoir of angular momentum in the stellar interior. This
reservoir must be large enough to spin up the surface of a giant and the cause
of the angular momentum dredge-up is unknown. Our results show that
some active stars do have larger abundances.
 However, we have only found a trend between rotation and
lithium excesses, and  if there is any clear connection (see next section),
 it should be very subtle
(such as a dependence in the rotational story as the component evolve off 
the MS and  ascent the RGB, very complex in close binaries due to the 
evolution of the convective envelope and the transport of angular momentum).
Therefore,  these data cannot discriminate
directly between these scenarios.

Some components with masses $\sim2.5$ M$_\odot$ have Li abundances close to the
cosmic value. 
This fact would indicate that, despite that they have expanded
and developed convective envelopes, they have not mixed the external material,
Li--rich, with the internal one, without it. Lithium could have been diffused
upwards during the pre-main sequence phase and would have achieved such high
concentrations in the surface Li-preservation zone that subsequent red giant
dilution decreased the Li only by a factor of 2--3 with respect to the cosmic
value. Since the expected dilution is a factor of about 50, the diffused
concentrations must be nearly a factor of 25 above cosmic. Such large surface
abundances (log N(Li)$\sim 4.5$) have never been reported in main sequence
stars, and thus this concentrated layer of Li must lie below the surface
convective zone but in the Li-preservation zone, a difficult theoretical feat.
An alternative explanation would be Li creation in the stellar interior. For
this mass range, Fekel \& Balachandran (1993) also found that a large fraction
of chromospherically active giants have high Li (See above). 
However, the  simplest explanation
is to assume that they have diluted partially the lithium due to the fact
 that the have just started to climb the RGB.

Some of the low mass stars ($\sim$1 M$_\odot$) have abundances larger
than the maximum values predicted by dilution, but under the average values
of Hyades MS stars (Log N(Li)=0.4--1.0 dex).
 Since most of the original Li is depleted in single stars during
the MS (e.g. Chaboyer et al. 1995),  either a little amount of Li has been
destroyed after the evolution off the MS, as happens in MS components of CABS
(Paper II), or a mechanism has prevented the
depletion during the present phase of the stellar evolution. 

Randich et al. (1993, 1994) found that a significant number of evolved  CABS
have  Li excesses with respect to the typical values observed in other evolved
stars of similar spectral types. They pointed out that the most plausible
interpretation was that the CABS have evolved from massive progenitors (M$>$1.5
M$_\odot$), who have suffered little or no depletion on the MS. However, they
did not find  the expected  correlation between the Li abundances and the
masses. In our case,  we have not found this trend either. Moreover, the spread
in the Li abundances for stars at the same mass range and the obvious Li
excesses in some stars with masses smaller than 1.5 M$_\odot$ are strong
indications that others mechanisms are acting, and inhibiting partially the Li
depletion for some CABS.

 From this diagram, we conclude that there are some
 indications of inhibition of the Li depletion as it
happens in MS components of CABS (Paper II). However, these
lithium excesses are very small, and the inhibition of the depletion seems
 to be only partial. Moreover, they are only present in a fraction
of the sample of evolved components of CABS.

%\section{Lithium, age and rotation}

\subsection{Lithium and age}

Using theoretical isochrones and the age--mass relationship, we have estimated
the ages of evolved components of CABS (Section 3.3).
 The comparison between these ages and
the Li abundances indicates that there is no relation between them. Figure 5
displays the   CABS  as circles (the size increases with the mass). As a
 comparison, we have included giant stars --solid diamonds--
belonging to several open clusters of different ages from Gilroy (1989).
 For any particular mass range, there is an
important dispersion of log~N(Li) and a lack of dependence with age.

In particular, stars having masses in the range 3.0 $>$ M/M$_\odot$ $>$ 2.0 
have a large spread in a very small age range. However, once the secondaries
(the less massive star in the system) and 6 Tri, whose mass is uncertain, are
removed, the spread is reduced  by a large amount, except for HR1023. Since
this group of stars has not depleted its initial Li content during the Pre-main
Sequence (PMS) or MS stages, the value should be around log~N(Li)=1.5 or less,
the maximum value permitted by dilution, as it happens with the observed
abundances.

In the case of the stars having masses in the range 1.2--1.6 M$_\odot$, the
dispersion could be explained due to the fact that these stars come from mid
F-type stars in the MS (the stars located in the Li gap, where the Li content
can decrease by 2 dex before the star evolves off the MS). The observed Li
dispersion hardly depends on rotation,  because although rotation inside the MS
could cause a dispersion, after the stars evolved off the MS this information
is lost due to the transport of angular momentum from the interior to the
surface. However, as shown in Figure 3a, these stars do have Li excesses.

Finally, stars with masses lower than 1.2 M$_\odot$ show abundances similar to
those characteristics of stars from open clusters. A very close look reveals
that evolved components of CABS with masses in the range 0.8--0.9 M$_\odot$
have Li abundances higher than equivalent stars member of open clusters, in the
same way of MS components of CABS (Paper II).

\subsection{Lithium and rotation}

Figure 6a shows Li abundance against the photometric period of CABS. Symbols
 are as in Figure 3a, and the size of the symbols increases with mass.
 There is a  trend between both quantities, although a 
large scatter is also present which does not disappear if a particular mass 
range is selected. In any case, it is obvious that there are no systems
 with large values of both photometric period and Li abundance. 

Observations of unevolved late-type stars in young open clusters like $\alpha$
Persei and the Pleiades indicate that there is a dependence of Li abundance on
rotation: the most rapid rotators are generally the most Li-rich objects and
exhibit a much narrower abundance dispersion than the slow rotators
(Garc\'{\i}a L\'opez et al. 1994; Randich et al. 1998). De Medeiros et al.
(1997) have found, analyzing 65 subgiants of spectral types F, G and K, that
there is a correlation between $v\sin i$ and log~N(Li), but only for single
stars. They also found discontinuities in both parameters for stars of spectral
type F8 IV --(B--V)$\sim$0.55--. Since the magnetic breaking seems to be
responsible for the rotational discontinuity, this work argues that  the
magnetic breaking can also produce the discontinuity in the Li abundance. In
our case, there are no such discontinuities. The orbit supplies angular
momentum to replace that one lost via magnetic breaking and we have not
observed any discontinuity in the $T_{\rm eff}$--log~N(Li) plane. Again,
rotation and evolution seem to be important factors in the Li depletion
phenomenon.

The particular way rotation influences the evolution of Li in evolved stars is
not clear. If transfer of angular momentum between the orbit and the rotation
inhibits the mixing process below the convective envelope, the Li abundance
should be larger in systems having short orbital periods, since P$_{\rm orb}$
would be coupled to the rotation due to tidal forces. However, several
asynchronous systems have large abundances (TY Pic, HR3385, 93 Leo, HR6469, AS
Cap) and, as a group, there is no difference in the behavior of the Li 
abundances between systems with and without synchronization (Figure 6b).
 On the other hand, there is a trend between the temperature and the
photometric period. Therefore, it is not clear the link between
rotation and Li abundances. As we pointed out before, Fekel \& Balachandran
(1993) have found that, in chromospherically active giants, high rotation is
connected with high Li abundance. 
 This mechanism is likely not acting in CABS since it seems to
appear only in AGB stars, more massive and in a more evolved evolutionary stage
than CABS.

Our interpretation of the high Li abundances of several CABS involves the
normal evolution of rotation and internal structure. When a star crosses the
giant gap, the core decreases its size while the convective envelope increases.
The conservation of angular momentum induces a shear effect between the core,
which rotates faster and faster, and the envelope, which is decreasing its
rotation rate. The effect would be a mix related to rotation which would cause
transport of Li well below the convective envelope and its annihilation.
Therefore, the Li abundances would be smaller than those predicted by dilution,
showing a scatter which depends on the initial angular momentum. In the case of
CABS, and specifically for those systems which have been synchronized during
the MS or when they have evolved off the MS (tidal forces depend on the size of
the convective envelope and stellar radius), the orbit acts as a source of
angular momentum to the rotation, avoiding the differential rotation between
the core and the convective envelope, and the turbulent mixing, keeping more Li
than single stars at the same evolutionary stage. If this mechanisms were
responsible for the observed abundances, the stellar rotational history
(including the evolution of orbital and rotational periods) would be very 
important in the way Li evolves with time but, unfortunately, this is an
information lost at this evolutionary stage.

\section{Summary and conclusions}

We have studied the lithium abundances of evolved components of CABS. As it
happens in late-type stars belonging to open clusters, there is a relation
between the Li abundance and  the effective temperature, although a very large
scatter is present. Since these stars have different masses (even in the case
of the same $T_{\rm eff}$), various mixing mechanisms should be working. 

We have found that some evolved components show partial inhibition of the Li
destruction, and there are several cases with abundances above the maximum
value predicted by dilution.  Part  of
these stars have Li abundances which do not depend on the age or the 
evolutionary stage. In particular, there is a very large scatter for those
stars having $\sim$5 M$_\odot$, which could indicate that the circularization 
process proposed for AGB stars or  creation of Li via the Cameron--Fowler
mechanism are  acting. They could also be due to the transport of material
associated with the transport of angular momentum, but our data cannot 
discriminate between them.

A Li abundance spread also appears in groups of the same age and mass. This
scatter is related to rotation, but the particular way  stellar rotation
influences the Li abundance in evolved components of these binaries is not
clear. However, the transfer of angular momentum  from the orbit to the
rotation in the Main Sequence and when the stars are evolving off it could
inhibit partially the mixing of material (and the dilution) and the destruction
of Li associated with it. However,  the problem is still unsolved and more work
is needed to understand  how Li evolves in low-mass evolved stars and, in
particular, in binary systems of different types.

\acknowledgements{
This research has made use of the Simbad database, operated at
CDS, Strasbourg,  France.
 We greatly appreciate the comments and suggestions on this paper
 by the referee, Dr. Sofia Randich, which helped to improve the presentation
and discussion of the results.
DBN acknowledges the support by the Universidad Complutense,   
the Real Colegio Complutense at Harvard University, and the
MEC/Fulbright commission. This work has been partially 
supported by the Spanish Direcci\'on General de Educaci\'on Superior (DGES) 
under projects PB92-0434-C02-01, PB94-0203, and PB95-1132-C02-01.}

\begin{center}
{\sc Figure Captions}
\end{center}

{Figure 1.-  Color-Magnitude diagram for our sample of evolved CABS. Note
that we have used the deconvolved photometry.}

{Figure 2.-  Several examples of the spectra.}

{Figure 3.- Lithium abundance against effective temperature. 
{\bf a} Giant components of CABS are shown as circles, and the size
 increases with stellar mass. 
{\bf b} Comparison between the CABS (solid circles)
and giant members of open clusters (open circles).
{\bf c} Comparison between CABS in the mass range
 1.1$\le$Mass$\le$1.5 M$_\odot$ (solid circles) and
 giants belonging to M67 (open circles).}

{Figure 4.- Lithium abundance against stellar mass.  The average Li abundances
of Hyades stars and the maximum abundance predicted by dilution 
are shown as a solid lines. The cosmic abundance appears as a dashed line. }

{Figure 5.- Li abundance against stellar age. Sizes of the circle
symbols increase with increasing 
stellar mass. Cluster giants are shown as solid diamonds.}

{Figure 6.- Relation between Li abundances and the photometric periods.
{\bf a} Size increases with stellar mass.
{\bf b} Synchronous and asynchronous systems (solid and
open circles, respectively).}

\end{document}